\documentclass{elsart}

\usepackage{graphics}
\usepackage{graphicx}
\usepackage{amssymb}
\usepackage{amsmath}
\usepackage{amsfonts}

\begin{document}

\title{Massive Primordial Black Holes \\ in Hybrid Inflation}

\author{S.~G. Rubin}

\address{Moscow State Engineering Physics
Institute, \\ Kashirskoe sh., 31, Moscow 115409, Russia\\ Center
for Cosmoparticle Physics "Cosmion", Moscow, 125047 Russia \\
E-mail: sergeirubin@list.ru}


\maketitle

\begin{abstract}Black hole formation in the framework of hybrid
inflation is considered. It is shown that this model of inflation
provides conditions for multiple black hole production.
\end{abstract}


It is well known that energy density fluctuations at the early Universe give
rise primordial black holes (BH) formation \cite{Novikov79e}. Widespread
opinion is that these BHs are small enough with masses somewhere in the range $
M_{BH}\sim 10^{-5}-10^{20}gram$ depending on specific model. What could be said
about BHs with masses in the range $10^{20}-10^{40}gram$? There exist a few
number of inflationary models that predict BH production at some conditions
\cite{Yoko98}, \cite{Ru28e}.

Here I discuss necessary conditions for BH formation in the
framework of hybrid inflation \cite{Linde91b}. It will be shown
that it is hard to avoid BH production at early stage of the
Universe evolution described by this model. New fluctuating
mechanism of BH formation elaborated in \cite{Ru28e} proved to be
significant one if a potential of the inflaton field possesses at
least two minima as it is in the case of hybrid inflation.

Potential of hybrid inflation has the form
\[
V(\chi ,\sigma )=\varkappa ^{2}\left( M^{2}-\chi ^{2}/4\right)
^{2}+\frac{ \lambda ^{2}}{4}\chi ^{2}\sigma
^{2}+\frac{1}{2}m^{2}\sigma ^{2}
\] Inflation takes place during slow rolling along the valley $\chi
=0,\sigma >\sigma _{c}$, see Figure.1. Black dot marks critical
point
 \[
\sigma _{c}=\sqrt{2}\frac{\varkappa }{\lambda }M.
\] When the field $\sigma $ passes this value, the motion along the
line $\chi =0,\sigma <\sigma _{c}$ becomes unstable and the field
$\chi $ quickly moves to one of the minima $\chi _{\pm }=\pm
2M,\sigma =0.$

The field motion is ruled by classical equations \begin{equation}
\ddot{\sigma}+3H\dot{\sigma}+\frac{1}{2}\lambda ^{2}\chi ^{2}\sigma
+m^{2}\sigma =0,  \label{ClMotion1}
\end{equation}
 \[
\ddot{\chi}+3H\dot{\chi}-\varkappa ^{2}\chi \left( M^{2}-\chi
^{2}/4\right) + \frac{1}{2}\lambda ^{2}\chi \sigma ^{2}=0,
\] were $H$ is Hubble parameter at this period.

Slow rolling, which is necessary condition for inflation, means
slow variation of the field $\sigma $ along the valley $\chi =0$
what takes place while $\sigma >\sigma _{c}$. This condition looks
like $m<<H$ for simple estimation. Neglecting second derivative as
usual, solution of Eq.~(\ref{ClMotion1}) acquires the form
\[
\sigma (t)=\sigma _{in}\exp \left( -\frac{m^{2}}{3H}t\right) ,\quad
m<<H.
\]

\begin{figure}
\includegraphics[angle=0,width=0.6\textwidth]{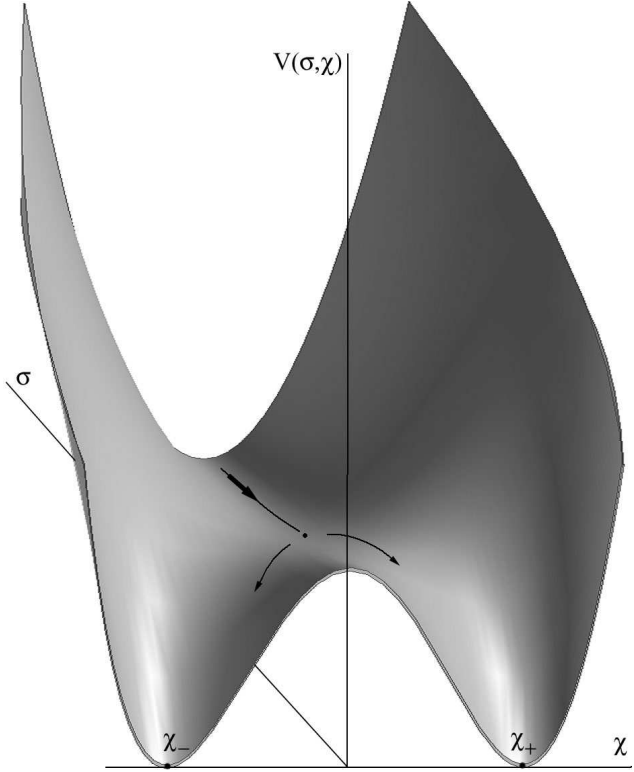}
\caption{Potential of Hybrid inflation. Arrows indicate the
directions of the classical motion of the inflaton.}
\label{fig8_1}
\end{figure}

The value of $\sigma _{in}$ could be obtained if we admit that the
period of inflation is about $N_{U}H^{-1}$ in terms of Hubble
parameter $H$. The value $N_{U}\approx 60$ is chosen for
estimations. Thus, from the condition $\sigma (N_{U}H^{-1})=\sigma
_{c}$ one obtains initial value of the field $\sigma $
\[
\sigma _{in}=\sigma_c\exp \left( \frac{m^{2}}{ 3H^{2}}N_{U}\right).
\]

Serious problem arises when we take into account field fluctuations
during inflation. Indeed, the classical motion along the line $\chi
=0$ describes an average motion. The fact is that the space is
constantly divided into increasing number of causally disconnected
domains. Each of them is characterized by a field value slightly
different from neighbor ones. Evidently, vast majority of them
contain field $\chi \neq 0$. Thus, at the end of inflation, i.e.
when the field $\sigma $ reaches the critical value $\sigma _{c}$
huge amount (about $ 10^{78}$) domains has been produced. Half of
them, those with $\chi <0,$ are directed to the minimum $\chi
_{-}=-2M,$ while the others -\ to the minimum $ \chi _{+}=+2M.$
After the inflation, we come to the Universe separated by
chaotically distributed domains with field values $\chi _{+}$ or
$\chi _{-}$ inside them. The neighboring domains are separated by a
field wall because a motion from $\chi _{+}$ to $\chi _{-}$ is
accompanied by crossing a space point with potential maximum at
($\chi =0,\sigma =0)$. Such a wall - dominated period is
unacceptable \cite{Zeldovich74}, because it prevents proper
evolution of the Universe. Consequently, motion along the value
$\chi =0$ , as it is usually supposed, is excluded.

The only way for our Universe to evolve into modern state is to be
created with initial field value $\chi _{in}\neq 0$ at the
beginning of inflation. During inflation, the field $\chi $ must
slowly approaches critical line $\chi =0$ for not to run into the
problem discussed above. One of the condition of slow motion for
the field $\chi $ is
\[
\eta \equiv \frac{V_{\chi \chi }^{\prime \prime }}{3H^{2}}<<1.
\] Estimation for this value can be easily performed for $\sigma
\approx \sigma _{in},\quad \chi =0$ and $H\approx \sqrt{8\pi
/3}\varkappa M/M_{P}$, $M_{P}$ is Plank mass. The result is
\begin{eqnarray*} \eta &\equiv &\frac{V_{\chi \chi }^{\prime \prime
}}{3H^{2}}\simeq \frac{ \frac{1}{2}\lambda ^{2}\sigma
_{in}^{2}-\varkappa ^{2}M^{2}}{3H^{2}} \\ &\approx &\frac{\varkappa
^{2}M^{2}\left( e^{\frac{2m^{2}}{3H^{2}} N_{U}}-1\right)
}{3H^{2}}\approx \frac{2N_{U}}{(8\pi )^{2}}\frac{
M_{P}^{4}m^{2}}{\varkappa ^{2}M^{6}},
\end{eqnarray*} what leads to inequality
\begin{equation}
\eta =\frac{6N_{U}}{(8\pi )^{2}}\frac{M_{P}^{4}m^{2}}{\varkappa
^{2}M^{6}} <<1.  \label{eta}
\end{equation}
 Combining it with the formula for temperature fluctuations
\cite{Lazarides99} , \cite{COBE}
\begin{equation}
\left( \frac{16\pi }{45}\right) ^{1/2}\frac{\lambda \varkappa
^{2}M^{5}}{ M_{P}^{3}m^{2}}\sim \frac{\delta T}{T}\approx 10^{-5},
\label{temper}
\end{equation} we obtain an estimation for the parameter $\lambda $ \begin{equation}
\lambda \approx \frac{(8\pi )^{2}}{6N_{U}}\sqrt{\frac{45}{16\pi
}}\frac{ \delta T}{T}\eta \frac{M}{M_{P}}.  \label{lam}
\end{equation} Evidently its numerical value is rather small unless $M>>M_{P}$.

It is worth to note that the hybrid model \cite{Linde91b} was
invented just to avoid the problem with very small coupling
constants. Meantime Eq.~(\ref {lam}) indicates unambiguously that
$\lambda <<1$ at reasonable values of parameters and hence this
problem remains in the hybrid inflation model.

If average field value approaches too close to critical line $\chi
=0,$ the fluctuations of the field in some space domains could
cross this line. In future, these domains will be filled by vacuum,
say, $\chi _{-}$ surrounded by a sea of another vacuum $\chi _{+}$.
The two vacua are separated by a closed wall as it was discussed
above. A number of such a walls depends in initial conditions at
the moment of our Universe creation.

Let us estimate energy and size of the closed walls. To proceed,
suppose that the field in the volume in question crossed critical
line at e-folds number $N$ before the end of inflation. Its size is
about Hubble radius, $ H^{-1}$ and it will be increased in $e^{N}$
times during inflation. Surface energy density of the domain wall
after inflation is
\begin{equation}
\epsilon =\frac{8\sqrt{2}}{3}\varkappa M^{3}.  \label{sigma}
\end{equation} Thus the energy $E_{wall}$ contained in the wall after inflation is
at least
\begin{equation}
E_{wall}\approx 4\pi \epsilon \left( H^{-1}e^{N}\right)
^{2}=4\sqrt{2}\frac{ M_{P}^{2}}{\varkappa M}e^{2N}.  \label{Ewall}
\end{equation} Numerical value $N$\ varies in the interval $\left( 0<N<N_{U}\approx
60\right) .$\

Gravitational radius of the wall could be easily calculated
\[
r_{g}=2E_{wall}/M_{P}^{2}\approx \frac{8\sqrt{2}}{\varkappa M}e^{2N},
\] what is much larger than the wall width $d=2\sqrt{2}/(\varkappa M)$
for any e-fold $N$. It means that this wall will collapse into BH
with mass $ M_{BH}\approx E_{wall}$ \cite{Ru28e}.

Let us estimate masses of such a BHs for ordinary values of the
parameters $ \varkappa =10^{-2}$ and $M=10^{16}GeV.$ If $N=40$ we
obtain the mass of BH
\[
M_{BH}\approx 3\cdot 10^{59}GeV\sim
100\hspace{1mm}Solar\hspace{1mm}mass.
\] The same estimation for a mass of smallest BHs which are created at
the e-fold number $N=1$ before the inflation is finished gives
 \[
M_{BH,small}\approx 10^{6}M_{P}.
\]

Thus, hybrid inflation leads to BH production in the wide range
$10^{25}\div 10^{59}$ GeV. A number of the massive Bhs depends on
how close average field value approaches to critical line. It, in
turn, depends on the initial conditions and specific values of
parameters of the model. Average dispersion of the field $\chi$ is
about
\[
\left\langle \delta \chi \right\rangle \approx \frac{H}{2\pi
}\sqrt{N_{U}}
\]
If the field $\chi $ approaches to this value during its classical
motion, overproduction of black holes is inevitable. Initial value
of the field $\chi $ must satisfy inequality
\begin{equation}
\chi _{in}\geq \frac{H}{2\pi }\sqrt{N_{U}}=\sqrt{\frac{2N_{U}}{3\pi
}}\frac{ \varkappa M^{2}}{M_{P}}
\end{equation}
what is necessary condition to avoid too many black holes after the
end of the inflation.

In conclusion, the mechanism of massive BH production revealed in
\cite{Ru28e} works effectively in the hybrid model of inflation. It
proves to be powerful tool for testing of inflationary models and
determining a range of their parameters. Careful investigation of
fluctuations in the framework of hybrid inflation indicates that
coupling constant must be very small to fit observations.

The author is grateful to M.Yu. Khlopov and A.S. Sakharov for
discussion.


\end{document}